\documentclass[3p,twocolumn,sort&compress]{elsarticle}
\usepackage{atlasphysics}
\usepackage{subfigure}
\usepackage{multirow}
\usepackage{amsmath}
\usepackage{amssymb}
\usepackage{hypernat}
\biboptions{numbers,square,comma,sort&compress}
\usepackage{hyperref}

\newcommand\htollllp{$H\to ZZ^{(*)}\to\ell^+\ell^-\ell^{'+}\ell^{'-}$}
\newcommand\htollllbrief{$H\to ZZ^{(*)}\to 4\ell$}

\newcommand\progname[1]{{\sc #1}}
\newcommand\pval{\ensuremath{p_0}}

\newcommand{\tabscript}[3]{%
 	\setlength{\fboxrule}{0pt}%
 	\fbox{\ensuremath{#1^{#2}_{#3}}}%
}
\newcommand\candmu{24~}
\newcommand\candmix{30~}
\newcommand\cande{17~}
\newcommand\candtotal{71~}
\newcommand\candmuexp{18.6~$\pm$~2.8~}
\newcommand\candmixexp{29.7~$\pm$~4.5~}
\newcommand\candeexp{13.4~$\pm$~2.0~}
\newcommand\bkgexp{62~$\pm$~9~}
\newcommand\lumiAverage{\ensuremath{4.8~\ifb}}
\newcommand\lumiFourMuon{\ensuremath{4.8~\ifb}}
\newcommand\lumiFourElectron{\ensuremath{4.9~\ifb}}
\newcommand\lumiTwoMuonTwoElectron{\ensuremath{4.8~\ifb}}

\newcommand\pvaluelowexp{10.6\%}
\newcommand\sigmalowexp{\ensuremath{1.3}}
\newcommand\pvaluelow{1.6\%}
\newcommand\sigmalow{\ensuremath{2.1}}
\newcommand\pvaluelowm{\ensuremath{125\gev}}
\newcommand\pvaluehighexp{0.14\%}
\newcommand\sigmahighexp{\ensuremath{3.0}}
\newcommand\pvaluehigh{1.3\%}
\newcommand\sigmahigh{\ensuremath{2.2}}
\newcommand\pvaluehighm{\ensuremath{244\gev}}
\newcommand\pvaluehightwoexp{7.1\%}
\newcommand\sigmahightwoexp{\ensuremath{1.5}}
\newcommand\pvaluehightwo{1.8\%}
\newcommand\sigmahightwo{\ensuremath{2.1}}
\newcommand\pvaluehighmtwo{\ensuremath{500\gev}}
\newcommand\excludedranges{$134-156\:\gev$, $182-233\:\gev$, $256-265\:\gev$ and $268-415\:\gev$}
\newcommand\excludedrangesexp{$136-157\:\gev$ and $184-400\:\gev$}
\newcommand\lumiAveragePLB{\ensuremath{2.1~\ifb}}
\journal{Physics Letters B}
\begin{document}
\begin{frontmatter}
\title{\vspace{-0.5cm} \flushright{\normalsize{CERN-PH-EP-2012-014}} \\ \center{Search for the Standard Model Higgs boson in the decay channel $H\ra ZZ^{(*)}\ra 4\ell$ with \lumiAverage\ of $pp$
collision data \\ at $\sqrt{s}=7$ TeV with ATLAS}} 
\author{The ATLAS Collaboration}
\address{}
\begin{abstract}
This Letter presents a search for the Standard Model Higgs boson in the decay channel \htollllp{}, where $\ell,\ell^{'}=e\text{ or }\mu$, 
using proton-proton collisions at $\sqrt{s}=7\:\tev$ recorded with the ATLAS detector and corresponding to an integrated luminosity of \lumiAverage. 
The four-lepton invariant mass distribution is compared with Standard Model background expectations to derive upper limits on the cross section of a Standard Model Higgs boson 
with a mass between $110\:\gev$ and $600\:\gev$. The mass ranges \excludedranges  ~are excluded at the 95\%  confidence level. 
The largest upward deviations from the background-only hypothesis are observed for  Higgs boson masses of $\pvaluelowm,\pvaluehighm\text{ and }\pvaluehighmtwo$ 
with local significances of \sigmalow, \sigmahigh~and \sigmahightwo~standard deviations, respectively. Once the look-elsewhere effect is considered, 
none of these excesses are significant.
\end{abstract}
\end{frontmatter}

\section{Introduction}
The search for the Standard Model (SM) Higgs boson~\cite{Englert:1964et,Higgs:1964pj,Guralnik:1964eu} is one of the most important aspects of the CERN Large Hadron Collider (LHC) physics programme. 
Direct searches performed at the CERN Large Electron-Positron Collider (LEP)  excluded at 95\% confidence level (CL) the production of a SM Higgs boson with mass, $m_H$, less than $114.4\:\gev$~\cite{Barate:2003sz}. 
The searches at the Fermilab Tevatron $p\bar{p}$ collider have excluded at 95\% CL the region $156<\mH < 177\:\gev$~\cite{tevhiggs}. At the LHC, results
from data collected in 2010 extended the search in the
region $200< m_H<600\:\gev$ by excluding a Higgs boson with cross section larger than $5-20$ times the SM prediction~\cite{Aad:2011qi,Chatrchyan:2011tz}. In ATLAS these results were extended further
using the first $1.04-2.28~\ifb$ of data recorded in 2011~\cite{Collaboration:2011ww,Aad:2011uq,Aad:2011va,Aad:2011ec,Collaboration:2011as,ATLAS:2011aa}. In particular, the $H\rightarrow
WW^{(*)}\rightarrow\ell^+\nu\ell^-\bar{\nu}$~search ~\cite{ATLAS:2011aa} excluded at 95\% CL the region $145 <\mH < 206\:\gev$. 

The search for the SM Higgs boson through the decay \htollllp{}, where $\ell,\ell^{'}=e\text{ or }\mu$, provides good sensitivity over a wide mass range. 
Previous results from ATLAS in this channel~\cite{Aad:2011uq} excluded three mass regions between 191~\GeV\ and 224~\GeV~at 95\% CL with a \lumiAveragePLB~data sample. 
This Letter presents an update of this search in the mass range from $110\:\gev$ to $600\:\gev$, superseding Ref.~\cite{Aad:2011uq}. 
Three distinct final states, $\mu^+\mu^-\mu^+\mu^-$~($4\mu$), $e^+e^-\mu^+\mu^-$~($2e2\mu$), and $e^+e^-e^+e^-$~($4e$), are selected. 
The largest background to this search comes from continuum $(Z^{(*)}/\gamma^*)(Z^{(*)}/\gamma^*)$ production, referred to as $ZZ^{(*)}$ hereafter. 
For $m_H<180\:\gev$, there are also important background contributions from $Z+\rm{jets}$ and $t\bar{t}$ production, 
where the additional charged lepton candidates arise either from decays of hadrons with $b$- or $c$-quark content or from misidentification of jets. 

The $\sqrt{s}=7\:\tev$ $pp$ collision data were recorded during 2011 with the ATLAS detector at the LHC and correspond to an integrated luminosity of
\lumiAverage ~\cite{Aad:2011dr,lumi2011}. This analysis is using more than twice the integrated luminosity of Ref.~\cite{Aad:2011uq}, 
including the data therein. 
The electron identification efficiency has been improved; 
furthermore the electron tracks have been refitted using a Gaussian-sum filter~\cite{GSF}, which corrects for energy losses due to bremsstrahlung. 
The analysis also benefits from recent significant improvements in the alignment of the inner detector and the muon spectrometer.
\newline
\newline

\section{The ATLAS Detector}
The ATLAS detector~\cite{Aad:2008zzm} is a multi-purpose particle physics detector with forward-backward symmetric cylindrical 
geometry\footnote{ATLAS uses a right-handed coordinate system with its origin at the nominal interaction point. 
The $z$-axis is along the beam pipe, the $x$-axis points to the centre of the LHC ring and the $y$-axis points upward. 
Cylindrical coordinates ($r$,$\phi$) are used in the transverse plane, $\phi$ being the azimuthal angle around the beam pipe. 
The pseudorapidity $\eta$ is defined as $\eta=-\ln[\tan(\theta/2)]$ where $\theta$ is the polar angle.}. 
The inner tracking detector (ID)~\cite{Aad:2010bx} covers $\left|\eta\right|<2.5$ and consists of a silicon pixel detector, 
a silicon microstrip detector, and a transition radiation tracker. 
The ID is surrounded by a thin superconducting solenoid providing a $2\:\rm{T}$ axial magnetic field. 
A high-granularity lead/liquid-argon (LAr) sampling calorimeter~\cite{:2009zp} measures the energy and the position of electromagnetic showers with $\left|\eta\right|<3.2$. 
LAr sampling calorimeters are also used to measure hadronic showers in the end-cap ($1.5<\left|\eta\right|<3.2$) and forward ($3.1<\left|\eta\right|<4.9$) regions, 
while an iron/scintillator tile calorimeter~\cite{:2010dd} measures hadronic showers in the central region ($\left|\eta\right|<1.7$). 
The muon spectrometer (MS)~\cite{:2010ha} surrounds the calorimeters and consists of three large superconducting air-core toroids, 
each with eight coils, a system of precision tracking chambers ($\left|\eta\right|<2.7$), and fast tracking chambers for triggering. 
A three-level trigger system~\cite{atlas:trigger} selects events to be recorded for offline analysis. 

\section{Data and Simulation Samples}
The data are subjected to quality requirements: events recorded during periods when the relevant detector components were not operating normally are rejected. 
The resulting integrated luminosity is \lumiFourMuon{}, \lumiTwoMuonTwoElectron{} and \lumiFourElectron{} for the $4\mu$, $2e2\mu$ and $4e$~final states, respectively.

The \htollllbrief{} signal is modelled using the \progname{powheg} Monte Carlo (MC) event generator~\cite{powheg4,powheg5}, which calculates separately the gluon-gluon and
vector-boson fusion production mechanisms with matrix elements up to next-to-leading order (NLO). The Higgs boson transverse momentum (\pt) spectrum in the gluon fusion process is reweighted to match
the calculation of Ref.~\cite{deFlorian:2011xf}, which includes quantum chromodynamics (QCD) corrections up to NLO and QCD soft-gluon resummations up to next-to-next-to-leading logarithm (NNLL). 
\progname{powheg} is interfaced to \progname{pythia}~\cite{pythia} for showering and hadronization, which in turn is interfaced to \progname{photos}~\cite{Golonka:2005pn} for quantum
electrodynamics (QED) radiative corrections in the final state and to \progname{tauola}~\cite{Jadach:1993hs,Golonka:2003xt} for the simulation of $\tau$ lepton decays. \progname{pythia} is used to simulate the production of a Higgs boson in association with a $W$ or a $Z$~boson.

The Higgs boson production cross sections and decay branching ratios~\cite{Djouadi:1997yw,Bredenstein:2006rh,Bredenstein:2006ha,Actis:2008ts}, as well as their uncertainties, are taken from Refs.~\cite{LHCHiggsCrossSectionWorkingGroup:2011ti,LHCHiggsCrossSectionWorkingGroup:2012vm}. 
The cross sections for the gluon fusion process have been calculated at next-to-leading order (NLO) in QCD~\cite{Djouadi:1991tka,Dawson:1990zj,Spira:1995rr}, 
and then at next-to-next-to-leading order (NNLO)~\cite{Harlander:2002wh,Anastasiou:2002yz,Ravindran:2003um}. 
In addition, QCD soft-gluon resummations up to NNLL are applied for the gluon fusion process~\cite{Catani:2003zt}. 
The NLO electroweak (EW) corrections are applied~\cite{Aglietti:2004nj,Actis:2008ug}. These results are compiled in Refs.~\cite{Anastasiou:2008tj,deFlorian:2009hc,Baglio:2010ae} assuming factorisation between QCD and EW corrections.
The cross sections for the vector-boson fusion process are calculated with full NLO QCD and EW corrections~\cite{Ciccolini:2007jr,Ciccolini:2007ec,Arnold:2008rz}, 
and approximate NNLO QCD corrections are available~\cite{Bolzoni:2010xr}.
The associated productions with a $W$ or $Z$ boson are calculated at NLO~\cite{Han:1991ia} and at NNLO~\cite{Brein:2003wg} in QCD, 
and NLO EW radiative corrections~\cite{Ciccolini:2003jy} are applied. 
The uncertainty in the production cross section due to the choice of QCD scale is $^{+12}_{-8}\%$ for the gluon fusion process, and
$\pm 1\%$ for the vector-boson fusion, associated $WH$ production, and associated $ZH$ production processes~\cite{LHCHiggsCrossSectionWorkingGroup:2011ti}. 
The uncertainty in the production cross section due to the parton distribution function (PDF) and $\alpha_{s}$ is $\pm 8\%$ for gluon-initiated process and $\pm 4\%$ for quark-initiated processes~\cite{Botje:2011sn,Alekhin:2011sk,Lai:2010vv,Martin:2009iq,Ball:2011mu}. 
The Higgs boson decay branching ratio to the four-lepton final state is predicted by \progname{prophecy4f}~\cite{Bredenstein:2006rh,Bredenstein:2006ha}, 
which includes the complete NLO QCD+EW corrections, 
interference effects between identical final-state fermions, and leading two-loop heavy Higgs boson corrections to the four-fermion width. Table~\ref{tab:MCSignal} gives the
production cross sections and branching ratios for \htollllbrief{} for several Higgs boson masses.   

The cross section calculations do not take into account the width of the Higgs~boson, which is implemented through a relativistic Breit-Wigner line shape applied at the event-generator level. 
It has been suggested~\cite{Seymour:1995qg,Passarino:2010qk,Anastasiou:2011pi,LHCHiggsCrossSectionWorkingGroup:2012vm} that effects related to off-shell Higgs~boson production and interference with other 
SM processes may become sizeable for the highest masses ($\mH>400\:\gev$) considered in this search.
In the absence of a full calculation, a conservative estimate of the possible size of such effects is included as a signal normalization systematic uncertainty following 
a parameterization as a function of $\mH$: $150\%\times\mH^3[\mathrm{TeV}]$, for $\mH\ge300\:\gev$~\cite{LHCHiggsCrossSectionWorkingGroup:2012vm}. 

\begin{table*}[!htb]
\centering
\caption{Higgs boson production cross sections for gluon fusion, vector-boson fusion and associated production with a $W$ or $Z$ boson in $pp$ collisions at $\sqrt{s}=7\:{\rm
TeV}$~\cite{LHCHiggsCrossSectionWorkingGroup:2011ti}. The quoted uncertainties correspond to the total theoretical systematic uncertainty. The production cross section for associated
production with a $W$ or $Z$ boson is negligibly small for $\mH>300\:\gev$. 
The decay branching ratio for $H \rightarrow 4\ell$, with $\ell = e\text{ or }\mu$, is reported in the last column~\cite{LHCHiggsCrossSectionWorkingGroup:2011ti}.\label{tab:MCSignal}}
\vspace{0.1cm}
{\footnotesize
\begin{tabular}{cccccc}
\hline\hline
$m_{H}$ & $\sigma\left(gg\to H\right)$ & $\sigma\left(qq'\to Hqq'\right)$ & $\sigma\left(q\bar{q}\to WH\right)$ & $\sigma\left(q\bar{q}\to ZH\right)$   & ${\rm BR}\left(H\to ZZ^{(*)}\to4\ell\right)$ \\
$[{\rm GeV}]$ & $[{\rm pb}]$ & $[{\rm pb}]$ & $[{\rm pb}]$ & $[{\rm pb}]$ & $[10^{-3}]$\\
\hline
130\rule[-1mm]{0mm}{4.7mm} & $14.1^{+2.7}_{-2.1}$& $1.154^{+0.032}_{-0.027}$ & $0.501\pm 0.020$          & $0.278\pm 0.014$ & $0.19$\\
150\rule[-1mm]{0mm}{4.7mm} & $10.5^{+2.0}_{-1.6}$& $0.962^{+0.028}_{-0.021}$ & $0.300\pm 0.012$          & $0.171\pm 0.009$ & $0.38$\\
200\rule[-1mm]{0mm}{4.7mm} & $5.2^{+0.9}_{-0.8}$ & $0.637^{+0.022}_{-0.015}$ & $0.103\pm 0.005$          & $0.061\pm 0.004$ & $1.15$\\
400\rule[-1mm]{0mm}{4.7mm} & $2.0\pm 0.3$        & $0.162^{+0.010}_{-0.005}$ & $-$                       & $-$              & $1.21$\\
600\rule[-1mm]{0mm}{4.7mm} & $0.33\pm0.06$       & $0.058^{+0.005}_{-0.002}$ & $-$                       & $-$              & $1.23$\\
\hline\hline
\end{tabular}
}
\end{table*}	
The $ZZ^{(*)}$ continuum background is modelled using \progname{pythia}. 
The \progname{mcfm}~\cite{Campbell:1999ah,Campbell:2011bn} prediction, including both quark-antiquark annihilation and gluon fusion at QCD NLO, 
is used for the inclusive total cross section and the shape of the invariant mass of the $ZZ^{(*)}$ system ($m_{ZZ^{(*)}}$). 
The QCD scale uncertainty has a $\pm 5\%$ effect on the expected $ZZ^{(*)}$ background, 
and the effect due to the PDF and $\alpha_{s}$ uncertainties is $\pm 4\%$ ($\pm 8\%$) for quark-initiated (gluon-initiated) processes. 
An additional theoretical uncertainty of $\pm 10\%$ on the inclusive $ZZ^{(*)}$ cross section is conservatively included 
due to the missing higher-order QCD corrections for the gluon-initiated process, 
and a correlated uncertainty on the predicted $m_{ZZ^{(*)}}$ spectrum is estimated by varying the gluon-initiated contribution by 100\%~\cite{Campbell:2009kg}.

The $Z+\rm{jets}$ production is modelled using \progname{alpgen}~\cite{Mangano:2002ea} and is divided into two sources: 
$Z+\rm{light~jets}$ ---~which includes  $Zc\bar{c}$ in the massless $c$-quark approximation and $Zb\bar{b}$ from parton showers~--- and 
$Zb\bar{b}$ using matrix element calculations that take into account the $b$-quark mass. 
The MLM~\cite{Mangano:2006rw} matching scheme is used to remove any double counting of identical jets produced via the matrix element calculation and the parton shower, 
but this scheme is not implemented for $b$-jets. 
Therefore, $b\bar{b}$ pairs with separation $\Delta R=\sqrt{\left(\Delta \phi\right)^2+\left(\Delta\eta\right)^2}>0.4$ 
between the $b$-quarks are taken from the matrix-element calculation, whereas for $\Delta R<0.4$ the parton-shower $b\bar{b}$~pairs are used. 
In this search the $Z+\rm{jets}$ background is normalized using control samples from data. 
For comparisons with simulation, the QCD NNLO \progname{fewz}~\cite{Melnikov:2006kv,Anastasiou:2003ds} and 
\progname{mcfm} cross section calculations are used for inclusive $Z$ boson and $Zb\bar{b}$ production, respectively. 
The $t\bar{t}$ background is modelled using \progname{mc@nlo}~\cite{Frixione:2003ei} and 
is normalized to the approximate NNLO cross section calculated using \progname{hathor}~\cite{Aliev:2010zk}. 
The effect of the QCD scale uncertainty on the cross section is $^{+4}_{-9}\%$, 
while the effect of PDF and $\alpha_{s}$ uncertainties is $\pm 7\%$. 
Both \progname{alpgen} and \progname{mc@nlo} are interfaced to \progname{herwig}~\cite{herwig} for parton shower hadronization and 
to \progname{jimmy}~\cite{jimmy} for the underlying event simulation. 

Generated events are fully simulated using the ATLAS detector simulation~\cite{simuAtlas} within the GEANT4 framework~\cite{GEANT4}. 
Additional $pp$ interactions in the same and nearby bunch crossings (pile-up) are included in the simulation. 
The MC samples are reweighted to reproduce the observed distribution of the mean number of interactions per bunch crossing in the data.

\begin{table*}[!hbt]
 \begin{center}
 \caption{Lower thresholds applied to $m_{34}$ for reference values of $m_{4\ell}$. 
   For $m_{4\ell}$ values between these reference values the selection requirement is obtained via linear interpolation.\label{tab:kincuts}}
 \vspace{0.1cm}
 \begin{tabular}{cccccccccc}
 \hline\hline
 $\tabscript{m}{}{4\ell}$ (\gev) & $\leq$120 & 130 & 140 & 150 & 160 & 165 & 180 & 190 & $\geq$200 \\
 $m_{34}$ threshold  (\gev) & 15   & 20  & 25 & 30 & 30 & 35 & 40 & 50 & 60 \\
 \hline\hline
 \end{tabular}
 \end{center}
\end{table*}	

\section{Lepton Identification and Event Selection}
The data considered in this analysis are selected using single-lepton or di-lepton triggers. 
For the single-muon trigger the \pt~threshold is $18\:\gev$, while for the single-electron trigger the transverse energy, \et, threshold is $20-22\:\gev$ depending on the LHC instantaneous luminosity. 
For the di-muon and di-electron triggers the thresholds are $\pt=10\:\gev$ for each of the muons, and $\et=12\:\gev$ for each of the electrons, respectively.

Electron candidates consist of clusters of energy deposited in the electromagnetic calorimeter that are associated to ID tracks. Electron tracks have been refitted using a Gaussian-sum filter. The electron candidates must satisfy 
a set of identification criteria~\cite{Aad:2011mk} that require the shower profiles to be consistent with those expected for electromagnetic showers and a well-reconstructed ID 
track pointing to the corresponding cluster. The electron transverse momentum is computed from the cluster energy and the track direction at the interaction point. 

\begin{figure*}[!htb]
  \centering 
   \subfigure[\label{fig:sub_higgs4mu}]{\includegraphics[width=0.45\linewidth]{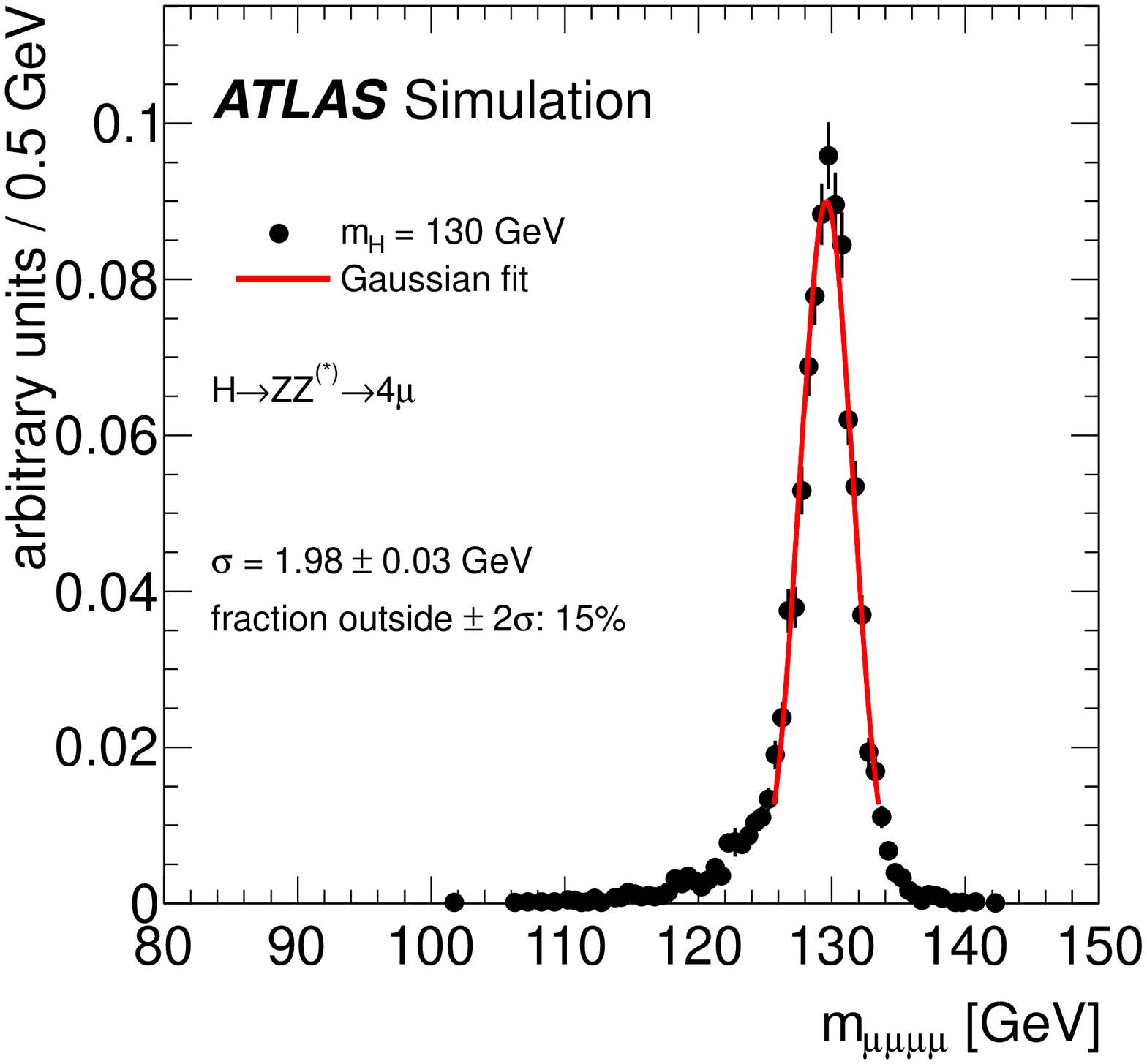}}
   \subfigure[\label{fig:sub_higgs4e}]{\includegraphics[width=0.45\linewidth]{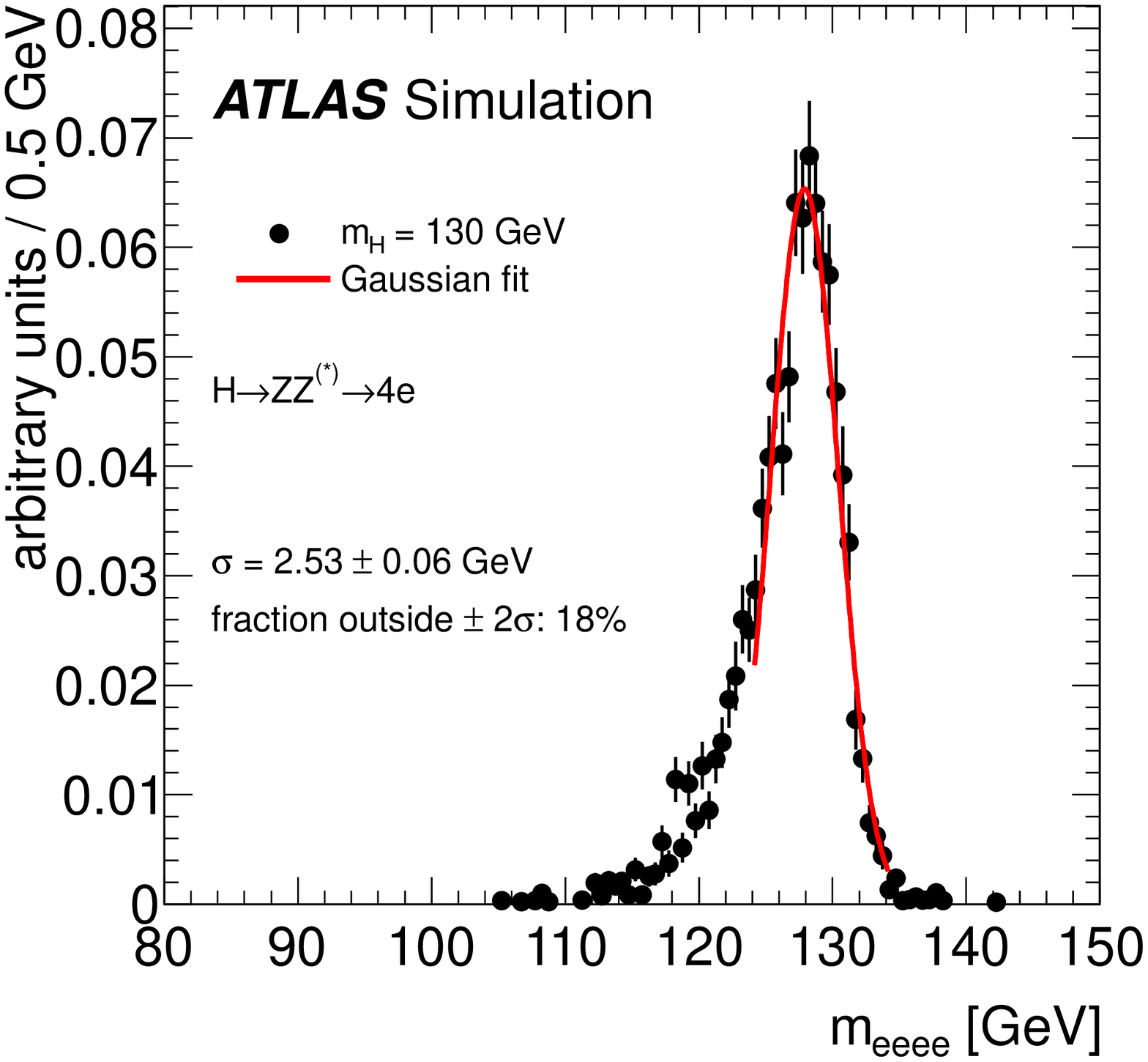}}
\caption{Invariant mass distributions for simulated \subref{fig:sub_higgs4mu} $H\rightarrow ZZ^{(*)}\rightarrow 4\mu$ and \subref{fig:sub_higgs4e} $H\rightarrow ZZ^{(*)}\rightarrow 4e$ events for
$m_{H}=130\:\gev$. The fitted range for the Gaussian is chosen to be : $-2~\sigma$ to $2~\sigma$ ($-1.5~\sigma$ to $2.5~\sigma$) for the $4\mu$ ($4e$) channel. The reduced mean value of the reconstructed invariant mass in the $4e$ channel arises from energy losses due to bremsstrahlung~\cite{Aad:2011mk}. The fraction of events outside the $\pm 2 \sigma$ region is found to be $15\%$ for $4\mu$ and $18\%$ for $4e$.\label{fig:higgs4l}}
\end{figure*}	  

Muon candidates are reconstructed by matching ID tracks with either complete or partial tracks reconstructed in the MS~\cite{Aad:2010yt}. 
If a complete track is present, the two independent momentum measurements are combined; otherwise the momentum is measured using the ID information only. 
To reject cosmic rays, muon tracks are required to have a transverse impact parameter, defined as the impact parameter in the transverse plane with respect to the primary vertex, of less than $1\:{\rm mm}$. 
The primary vertex is defined as the reconstructed vertex with the highest $\sum\pt^2$ of associated tracks among the reconstructed vertices with at least three associated tracks.

This analysis searches for Higgs boson candidates by selecting two same-flavour, opposite-sign lepton pairs in an event. 
The impact parameter of the leptons along the beam axis is required to be within $10\:{\rm mm}$ of the primary vertex. 
Each lepton must satisfy $\pt>7\:\gev$ and be measured in the pseudorapidity range $\left|\eta \right|<2.47$ for electrons and $\left|\eta\right|<2.7$ for muons. 
At least two leptons in the quadruplet must have $\pt >20\:\gev$. 
The leptons are required to be separated from each other by $\Delta R>0.1$. 
The invariant mass of the same-flavour and opposite-sign lepton pair closest to the $Z$ boson mass ($m_{Z}$) 
is denoted by $m_{12}$ and $\left|m_{Z}-m_{12}\right|<15\:\gev$ is required. 
The invariant mass of the remaining same-flavour and opposite-sign lepton pair, $m_{34}$, is required to be in the range $m_{\rm{min}} < m_{34} < 115\:\gev$, 
where $m_{\rm{min}}$ depends on the reconstructed four-lepton invariant mass, $m_{4\ell}$, as shown in Table~\ref{tab:kincuts}. 

The $Z+\rm{jets}$ and $t\bar{t}$ background contributions are further reduced by applying track- and calorimeter-based isolation and impact parameter requirements on the leptons.
For a lepton to be isolated, the sum of the \pt~of tracks within $\Delta R<0.2$ of the lepton divided by the lepton \pt~is required to be less than 0.15, 
while the sum of the \et~of the calorimeter cells with $\Delta R<0.2$ around the lepton divided by the lepton \pt~is required to be less than 0.3. 
The lepton track and the energies of calorimeter cells associated to it are excluded from the sum. 
Any contributions arising from other leptons of the quadruplet are subtracted. 
To reduce the impact of event pile-up, the tracks included in the \pt~sum for track isolation must be associated with the primary vertex, 
and the transverse energy included in the \et~sum for calorimeter isolation is corrected by subtracting 
a small amount of energy that depends on the number of reconstructed vertices in the event. 
In events with four-lepton invariant mass ($m_{4\ell}$) below $190\:\gev$, the transverse impact parameter significance, 
defined as the transverse impact parameter divided by the corresponding uncertainty, 
for the two lowest \pt~leptons in the quadruplet is required to be less than 3.5 (6) for muons (electrons).

The combined signal reconstruction and selection efficiencies for $\mH=130\:\gev$ ($\mH=360\:\gev$) are $27$\% ($60$\%) for the $4\mu$ channel, 
$18$\% ($52$\%) for the $2e2\mu$ channel and $14$\% ($45$\%) for the $4e$ channel. 
The final discriminating variable is $m_{4\ell}$, for which Higgs boson production would appear as a clustering of events. In Fig. \ref{fig:higgs4l}, the invariant mass
distributions for the $4\mu$ and $4e$ channels are presented for a simulated signal sample with $m_{H}=130\:\gev$. 
The width of the reconstructed Higgs boson mass distribution is dominated by experimental resolution for $m_{H}<350\:\gev$, 
while for higher $m_{H}$ the reconstructed width is dominated by the natural width of the Higgs boson; 
the predicted full-width at half-maximum is approximately $35\:\gev$ at $m_{H} = 400\:\gev$. 

\begin{figure*}[!htb]
\centering
\subfigure[\label{fig:sub_2crmu}]{\includegraphics[width=0.45\linewidth]{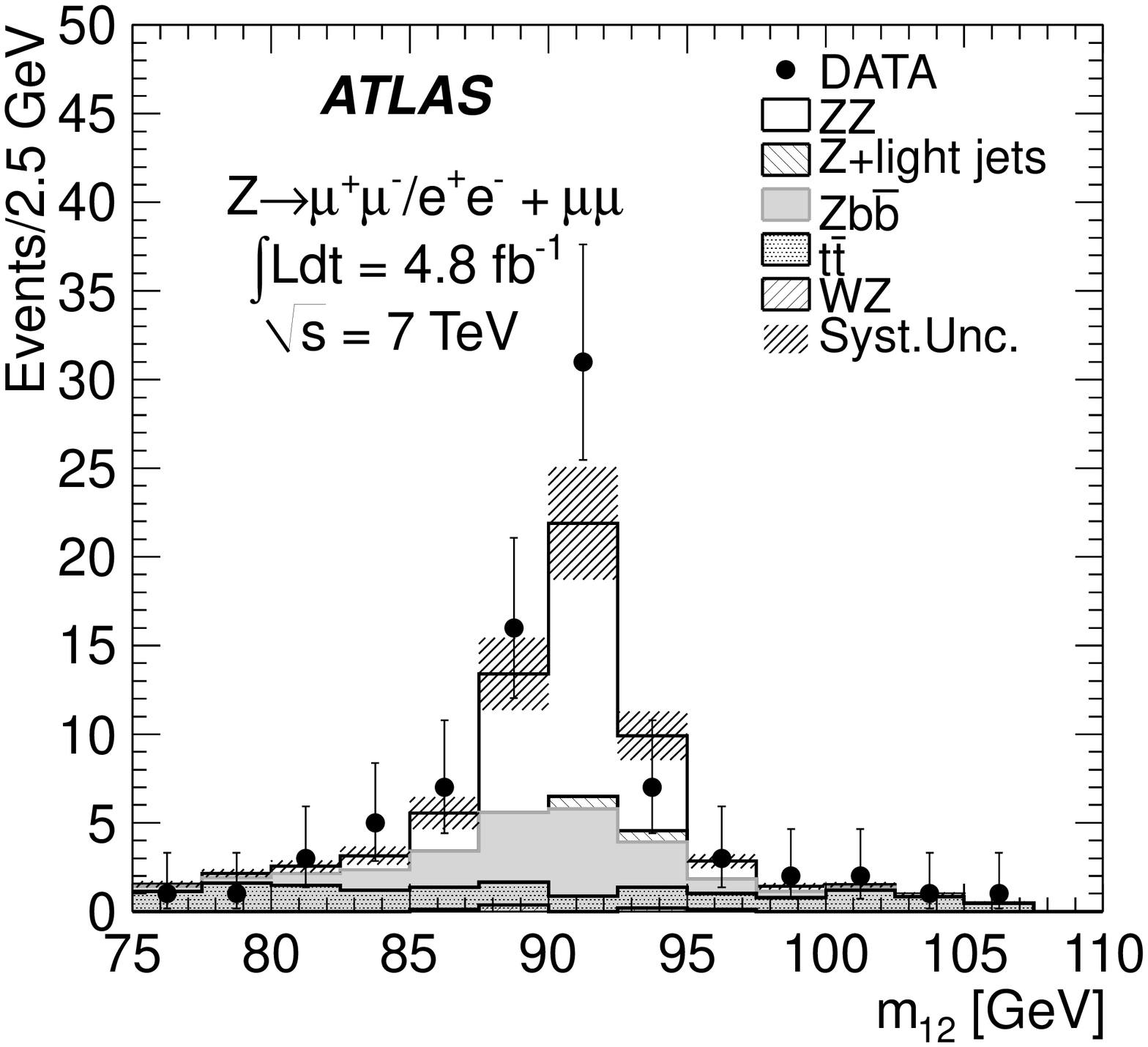}}
\subfigure[\label{fig:sub_2cre}]{\includegraphics[width=0.45\linewidth]{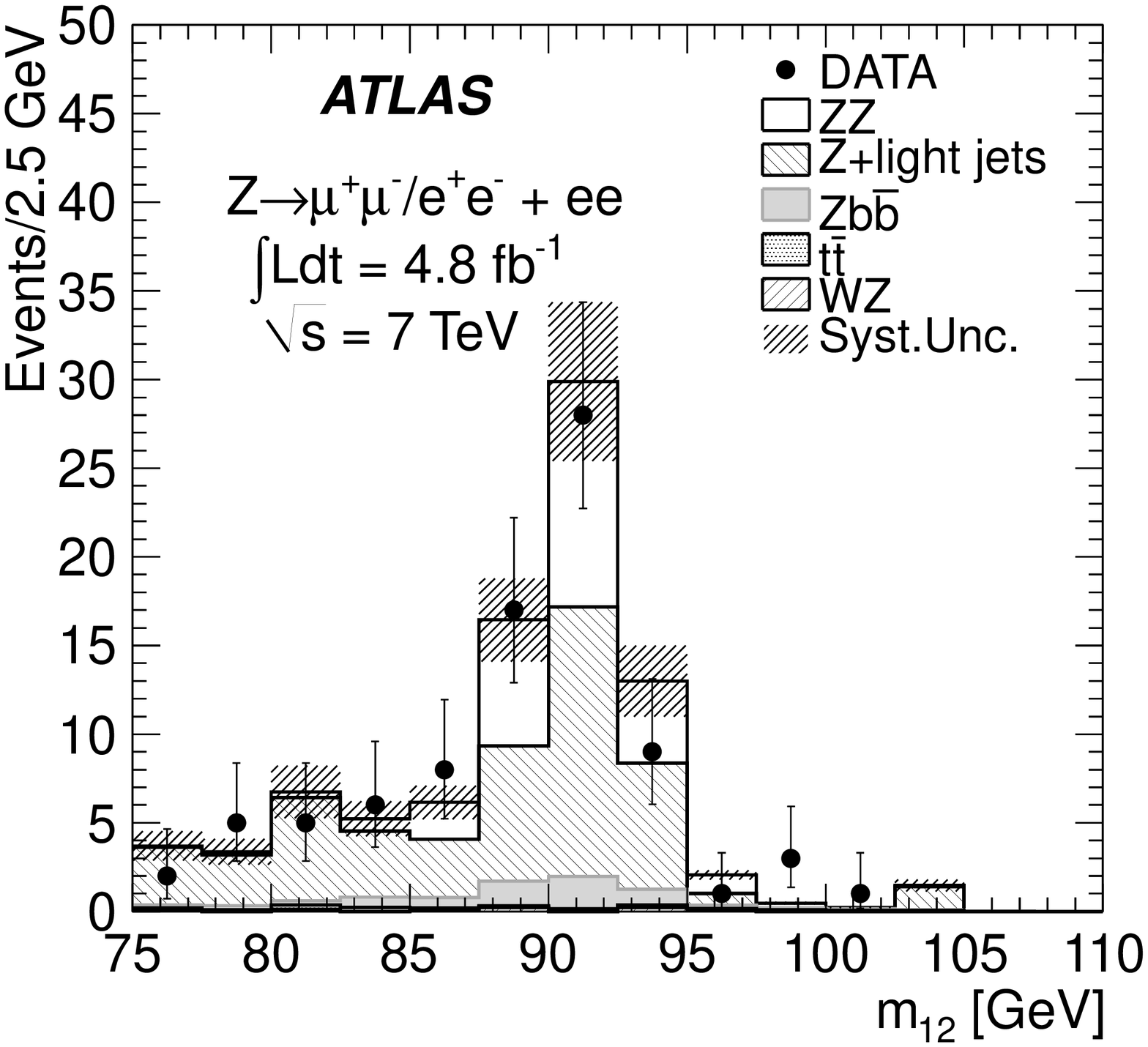}}\\
\subfigure[\label{fig:sub_crmu}]{\includegraphics[width=0.45\linewidth]{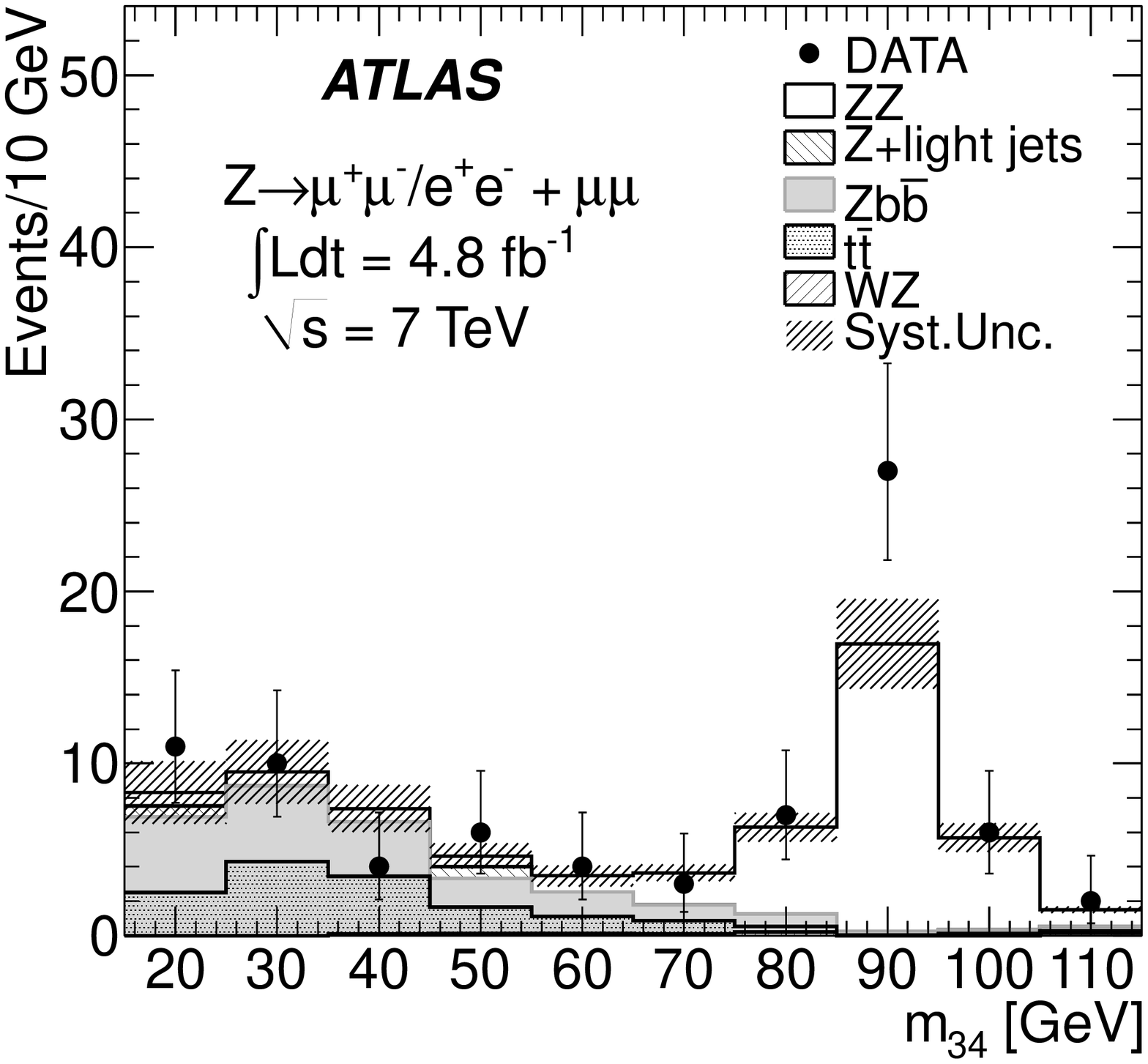}}
\subfigure[\label{fig:sub_cre}]{\includegraphics[width=0.45\linewidth]{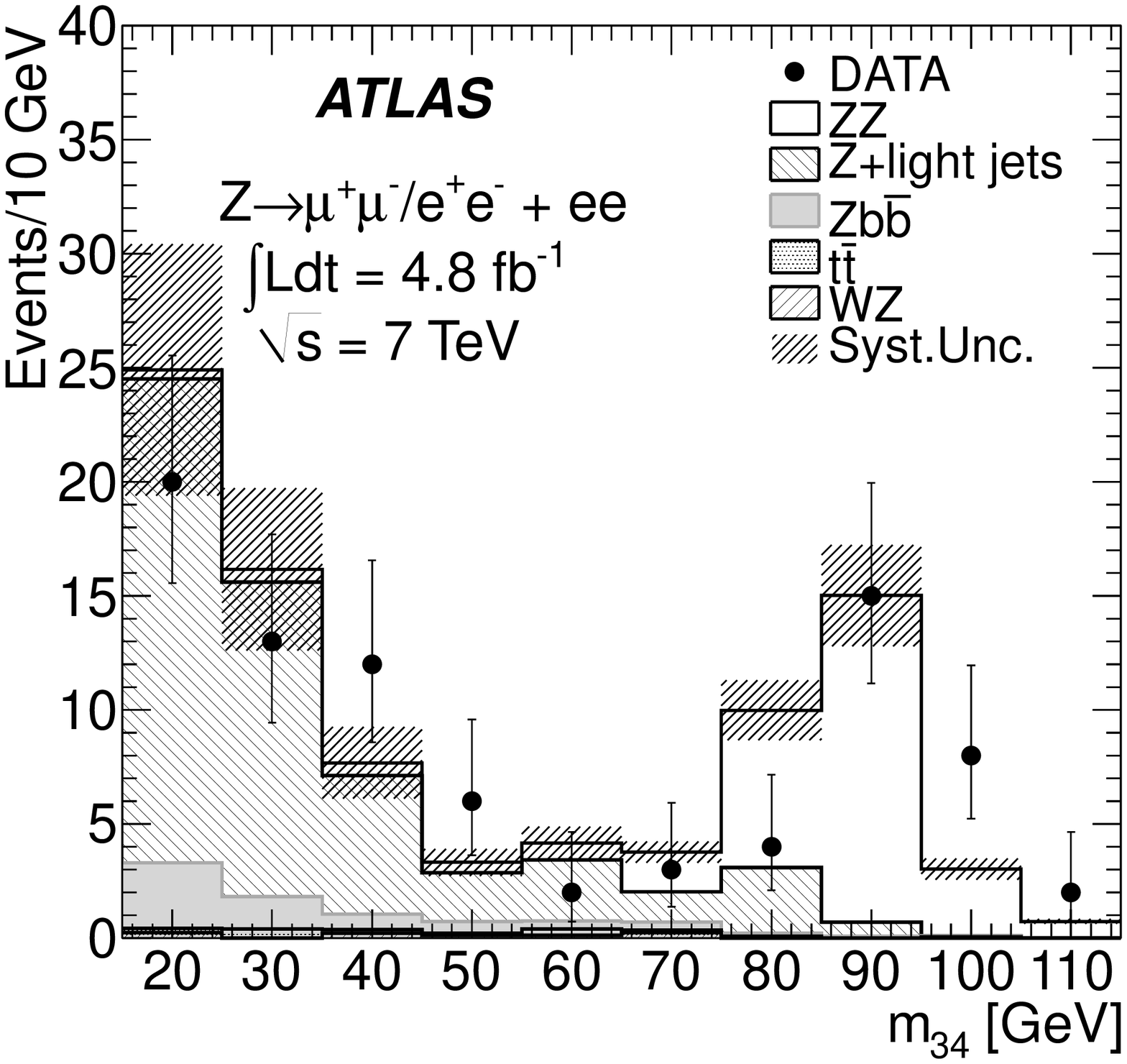}}
\caption{Invariant mass distributions of the lepton pairs in the control sample defined by a $Z$ boson candidate and an additional same-flavour lepton pair. 
The sample is divided according to the flavour of the additional lepton pair. 
In \subref{fig:sub_2crmu} the $m_{12}$ and in \subref{fig:sub_crmu} the $m_{34}$ distributions are presented for $Z(\rightarrow \mu^+\mu^-/e^+e^-)+\mu\mu$ events.
In \subref{fig:sub_2cre} the $m_{12}$ and in \subref{fig:sub_cre} the $m_{34}$ distributions are presented for $Z(\rightarrow \mu^+\mu^-/e^+e^-)+ee$ events.
The kinematic selections of the analysis are applied. Isolation requirements are applied to the first lepton pair only.\label{fig:sub_cr}}
\end{figure*}

\section{Background Estimation}
The expected background yield and its composition is estimated using MC simulation normalized to the
theoretical cross section for $ZZ^{(*)}$ production and by data-driven methods for the $Z+\rm{jets}$ and $t\bar{t}$ processes.

A control sample consisting of  $Z\to \ell^+\ell^-$ candidates with an additional loosely selected ---~no isolation or impact parameter requirements~--- same-flavour lepton pair 
is used to study the contributions of $Zb\bar{b}$ and $Z+\rm{light~jets}$. 
The $Zb\bar{b}$ background dominates the $Z+\rm{\mu\mu}$ sample, and the $Z+\rm{light~jets}$ background dominates in the $Z+ee$ sample.
The heavy flavour contribution in the $Z+\mu\mu$ control sample is estimated by subtracting from the data the light jet component. 
The latter is obtained in a data-driven manner by using measurements of the rate at which other particles are misidentified as muons. 
The $Z+\rm{light~jets}$ contribution in the $Z+ee$ final state is estimated by extrapolation, using MC simulation, 
from a background-dominated region defined by inverting the electron identification requirement on the transverse shower shape of the electromagnetic energy deposit.
These data-driven backgrounds are extrapolated to the signal region by applying the efficiencies found in MC simulation, and verified using data, 
for the isolation and impact parameter significance requirements.

The normalization of the $t\bar{t}$ background, which also contributes substantially in the $Z+\mu\mu$ 
final state, is verified using a control region of events containing an opposite-sign electron-muon pair consistent 
with the $Z$ boson mass and two additional same-flavour leptons. 

Figure~\ref{fig:sub_cr} displays the invariant masses of lepton pairs in events with a $Z$ boson candidate and an additional same-flavour lepton pair, 
selected by following the kinematic requirements of the analysis, and by applying isolation requirements to the first lepton pair only. 
The events are divided according to the flavour of the additional lepton pair into $Z+\mu\mu$ and $Z+ee$ samples, where $Z\rightarrow \mu^+\mu^-/e^+e^-$.
In Figs.~\ref{fig:sub_2crmu} and \ref{fig:sub_crmu} the $m_{12}$ and $m_{34}$ distributions are presented for $Z+\mu\mu$ events, 
while in Figs.~\ref{fig:sub_2cre} and \ref{fig:sub_cre} the corresponding distributions are presented for $Z+ee$ events.
The shapes and normalizations of the backgrounds discussed earlier are in good agreement with data; 
this is observed both for large values of $m_{34}$, where the $ZZ^{(*)}$ background dominates, and for low $m_{34}$ values.

\begin{table*}[!htb]
\begin{center}
\caption{The expected numbers of background events, with their systematic uncertainty, separated into ``Low-$m_{4\ell}$'' ($m_{4\ell}<180\:\gev$) and ``High-$m_{4\ell}$'' ($m_{4\ell}\geq 180\:\gev$)
regions, compared to the observed numbers of events. The expectations for a Higgs boson signal for five different $m_H$ values are also given.\label{tab:EventYields}}
\begin{tabular}{ccccccc}
\hline\hline
  & \multicolumn{2}{c}{$\mu^+\mu^-\mu^+\mu^-$}& \multicolumn{2}{c}{$e^+e^-\mu^+\mu^-$} & \multicolumn{2}{c}{$e^+e^-e^+e^-$} \\
  & Low-$m_{4\ell}$ & High-$m_{4\ell}$& Low-$m_{4\ell}$ & High-$m_{4\ell}$& Low-$m_{4\ell}$ & High-$m_{4\ell}$\\
 \hline
 Int. Luminosity  & \multicolumn{2}{c}{ \lumiFourMuon{}}& \multicolumn{2}{c}{\lumiTwoMuonTwoElectron{}} & \multicolumn{2}{c}{\lumiFourElectron{}} \\
 \hline
  $\tabscript{ZZ}{(*)}{}$          & 2.1 $\pm$ 0.3    & 16.3 $\pm$ 2.4    & 2.8 $\pm$ 0.6  & 25.2 $\pm$ 3.8   & 1.2 $\pm$ 0.3   & 10.4 $\pm$ 1.5  \\
  $Z+\rm{~jets}$ and $t\bar{t}$    & 0.16 $\pm$ 0.06  & 0.02 $\pm$ 0.01   & 1.4 $\pm$ 0.5  & 0.17 $\pm$ 0.08  & 1.6 $\pm$ 0.7   & 0.18 $\pm$ 0.08\\\hline
  Total Background                 & 2.2 $\pm$ 0.3    & 16.3 $\pm$ 2.4    & 4.3 $\pm$ 0.8  & 25.4 $\pm$ 3.8   & 2.8 $\pm$ 0.8   & 10.6 $\pm$ 1.5\\\hline 
  Data                             & 3              & 21              & 3            & 27             & 2             & 15\\
  \hline 
  $m_{H}=130\:\gev$ &\multicolumn{2}{c}{1.00 $\pm$ 0.17} &\multicolumn{2}{c}{1.22 $\pm$ 0.21}&\multicolumn{2}{c}{0.43 $\pm$ 0.08}\\
  $m_{H}=150\:\gev$ &\multicolumn{2}{c}{2.1 $\pm$ 0.4} &\multicolumn{2}{c}{2.9 $\pm$ 0.4}&\multicolumn{2}{c}{1.12 $\pm$ 0.18}\\
  $m_{H}=200\:\gev$ &\multicolumn{2}{c}{4.9 $\pm$ 0.7} &\multicolumn{2}{c}{7.7 $\pm$ 1.0}&\multicolumn{2}{c}{3.1 $\pm$ 0.4} \\
  $m_{H}=400\:\gev$ &\multicolumn{2}{c}{2.0 $\pm$ 0.3} &\multicolumn{2}{c}{3.3 $\pm$ 0.5}&\multicolumn{2}{c}{1.49 $\pm$ 0.21} \\
  $m_{H}=600\:\gev$ &\multicolumn{2}{c}{0.34 $\pm$ 0.04}&\multicolumn{2}{c}{0.62 $\pm$ 0.10}&\multicolumn{2}{c}{0.30 $\pm$ 0.06} \\
\hline\hline
\end{tabular}
\end{center}
\end{table*}       

\begin{figure*}[!htb]
  \centering 
  \subfigure[\label{fig:sub_m12data}]{\includegraphics[width=0.45\linewidth]{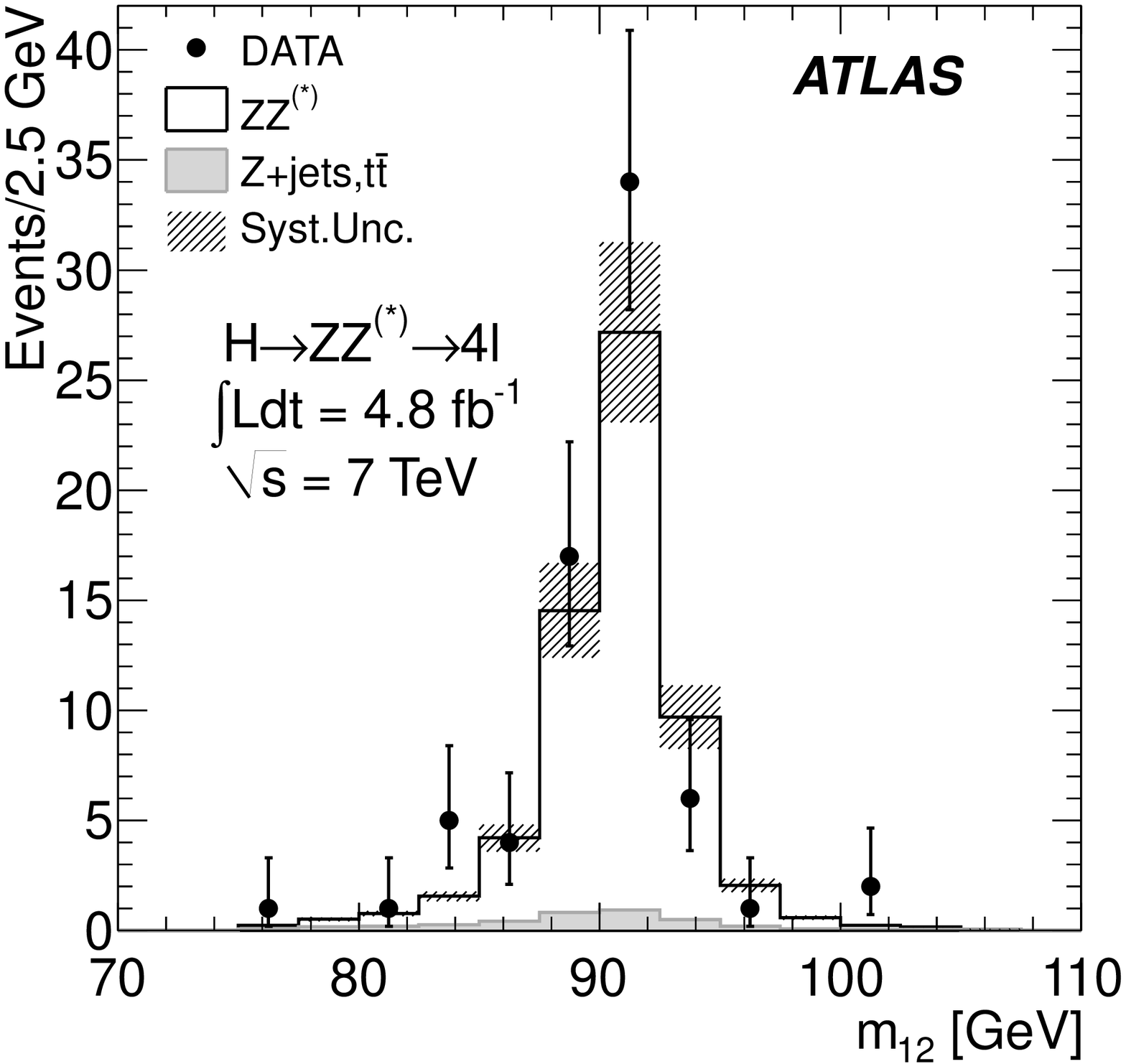}}
  \subfigure[\label{fig:sub_m34data}]{\includegraphics[width=0.45\linewidth]{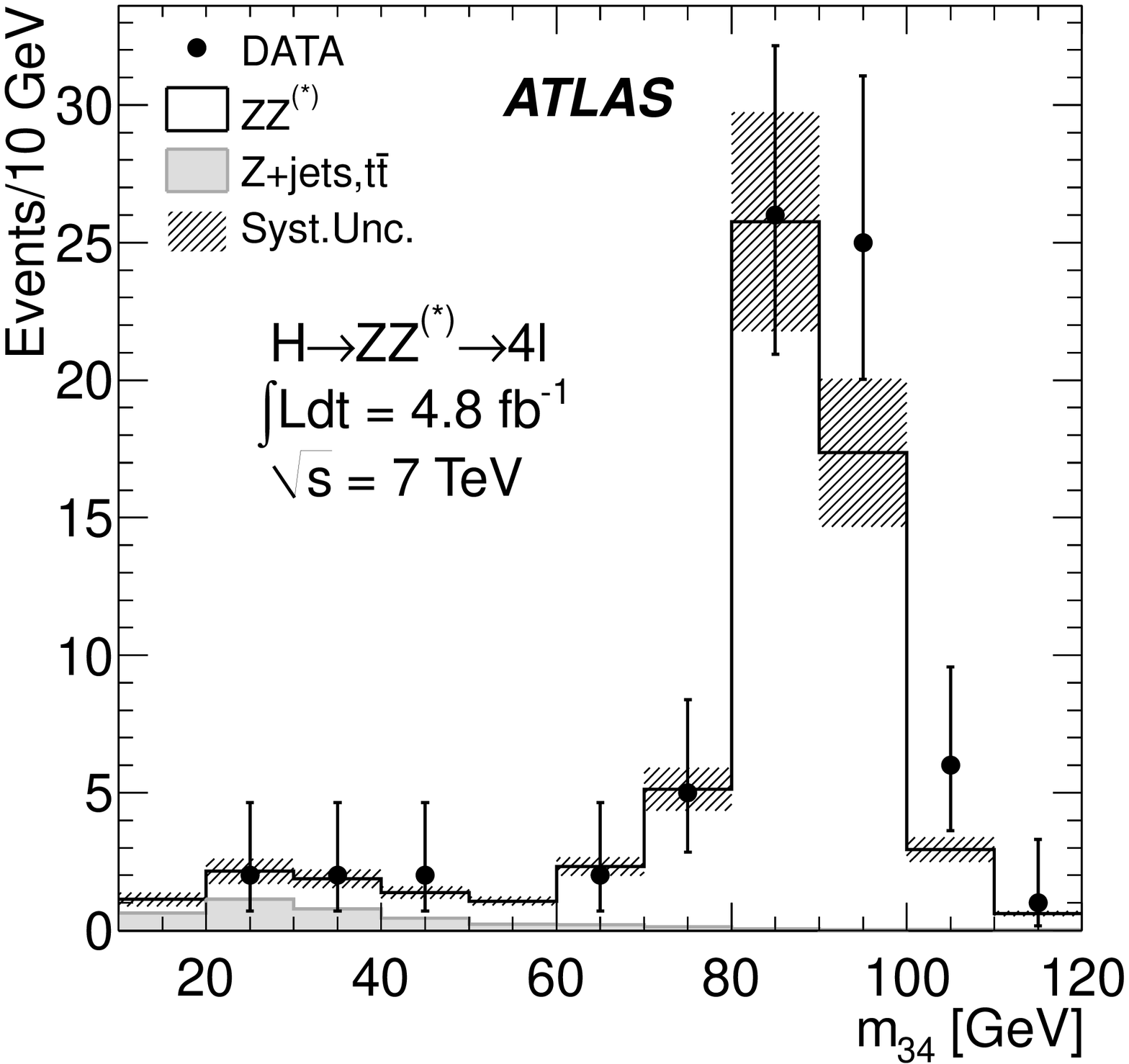}}
\caption{Invariant mass distributions \subref{fig:sub_m12data} $m_{12}$ and \subref{fig:sub_m34data} $m_{34}$ for the selected candidates. The data (dots) are compared to the background expectations from the dominant $ZZ^{(*)}$ process and the sum of $t\bar{t}$, $Zb\bar{b}$ and $Z+\rm{light~jets}$ processes. Error bars represent 68.3\% central confidence intervals.\label{fig:finalMasses}}
\end{figure*}  

\section{Systematic Uncertainties}
Uncertainties in lepton reconstruction and identification efficiency, and on the momentum resolution and scale, 
are determined using samples of $W$, $Z$ and $J/\psi$ decays. 
The muon efficiency uncertainty results in a relative acceptance uncertainty in the signal and the $ZZ^{(*)}$ background 
which is uniform over the mass range of interest, and amounts to 0.22\% (0.16\%) for the $4\mu$ ($2e2\mu$) channel. 
The uncertainty in the electron efficiency results in a relative acceptance uncertainty of 2.3\% (1.6\%) 
for the $4e$ ($2e2\mu$) channel at $m_{4\ell}=600\:\gev$ and reaches 8.0\% (4.1\%) at $m_{4\ell}=110\:\gev$. 
The effects of muon momentum resolution and scale uncertainties are found to be negligible. 
The energy resolution uncertainty for electrons is negligible, 
while the electron energy scale uncertainty results in an uncertainty of less than 0.6\% (0.3\%) on the mass scale of the $m_{4\ell}$ distribution for the $4e$($2e2\mu$) channel. 

The selection efficiencies of the isolation and impact parameter requirements are studied using data for both isolated and non-isolated leptons. 
Isolated leptons are obtained from $Z\rightarrow\ell\ell$ decays, while additional leptons reconstructed in events with $Z\rightarrow\ell\ell$ 
decays constitute the sample of non-isolated leptons. 
Additional checks are performed with non-isolated leptons from semi-leptonic $b$- and $c$-quark decays in a heavy-flavour enriched di-jet sample. 
Good agreement is observed between data and simulation and the systematic uncertainty is, in general, estimated to be small with respect to the other systematic uncertainties. 
An exception is found in the case of isolated electrons with $\et<15\:\gev$, 
where due to the small number of $Z\rightarrow e^+e^-$ events and the substantial QCD backgrounds an additional uncertainty of 5\% is added.

An additional uncertainty in the signal selection efficiency is added due to the modelling of the signal kinematics. 
This is evaluated by varying the Higgs boson \pt~spectrum in the gluon fusion process according to the PDF and QCD scale uncertainties.

\begin{figure*}[!htb]
  \centering
    \subfigure[\label{fig:sub_zoom_range}]{\includegraphics[width=0.45\linewidth]{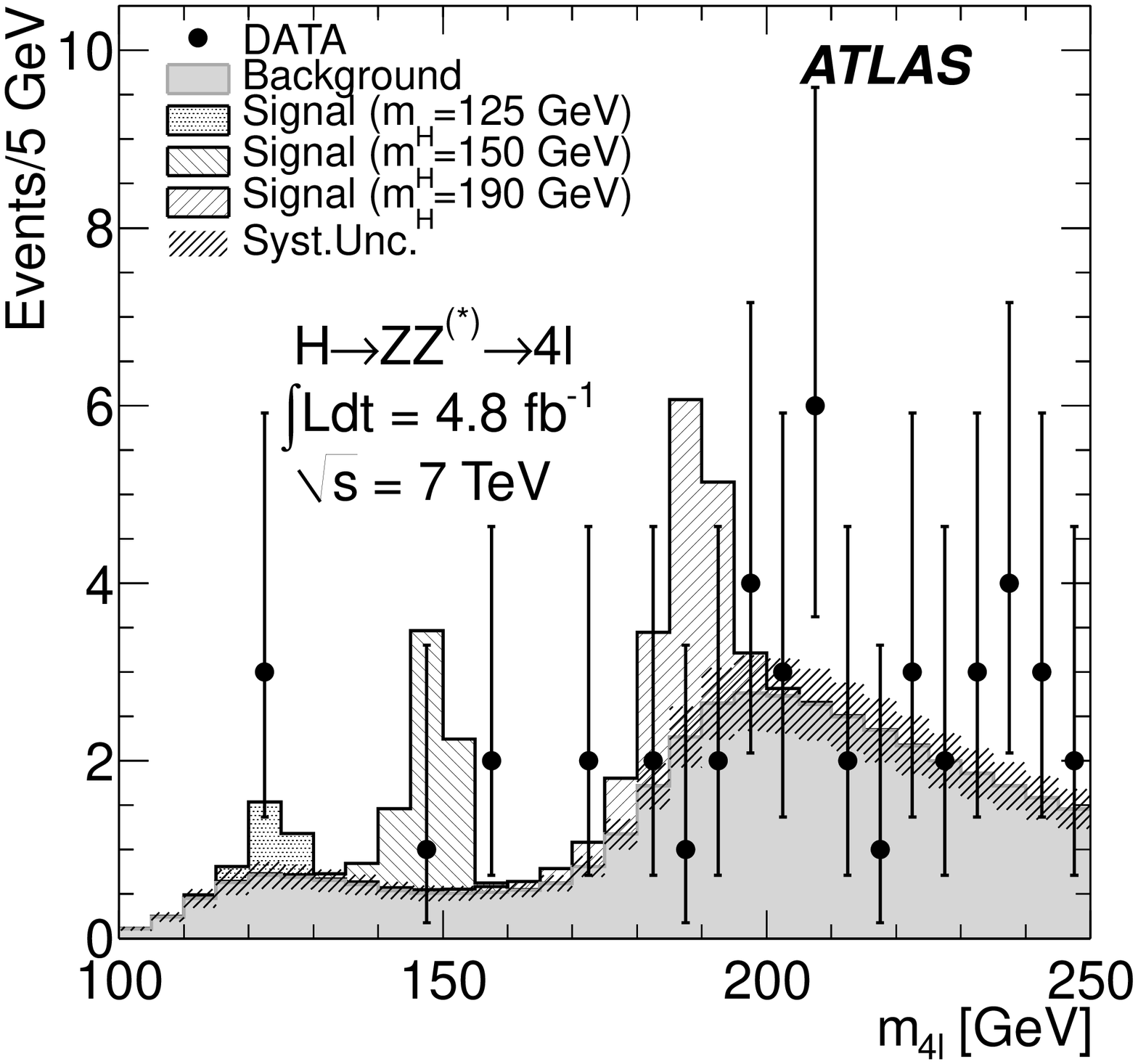}}
    \subfigure[\label{fig:sub_full_range}]{\includegraphics[width=0.45\linewidth]{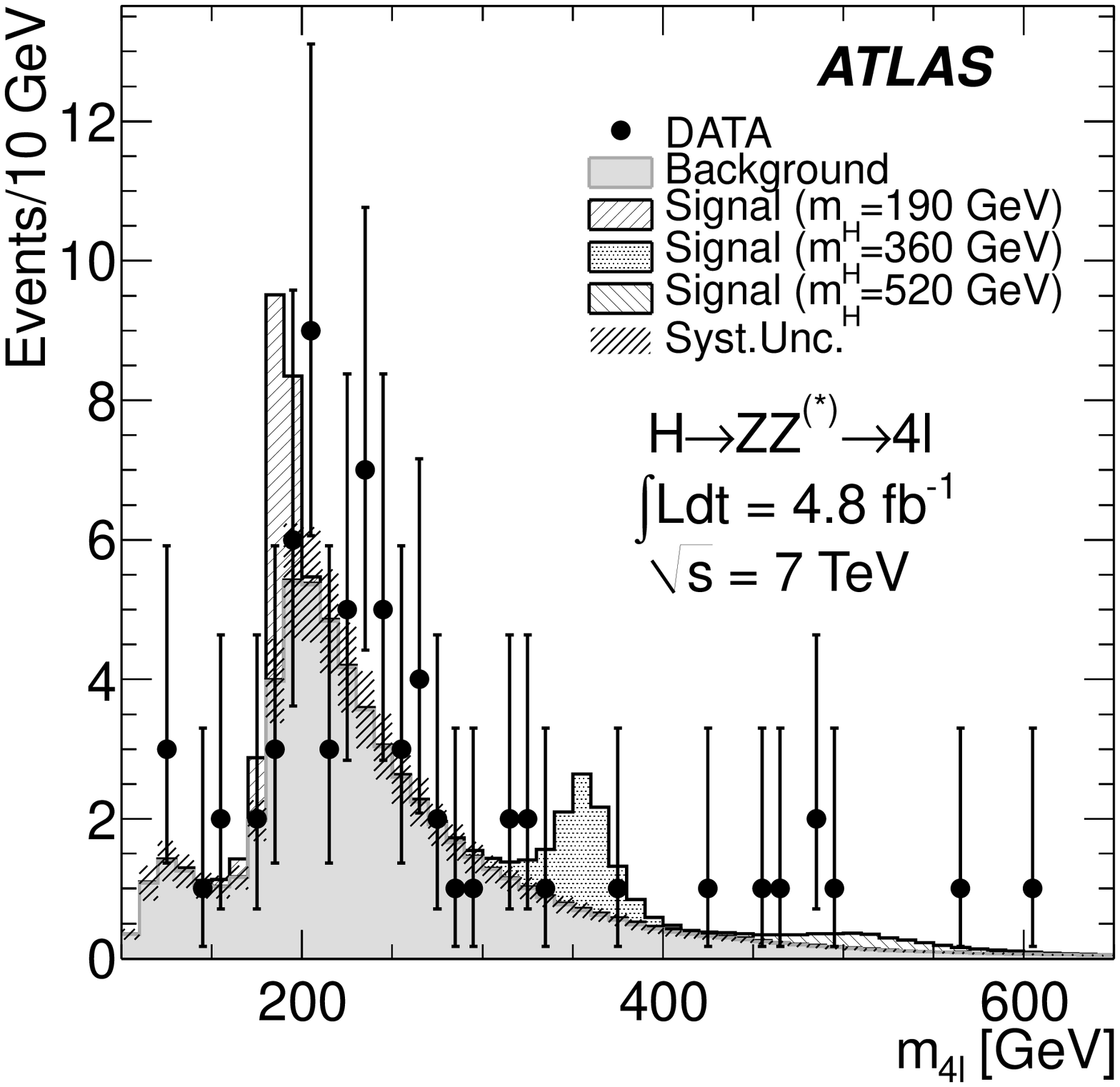}}
  \caption{$m_{4\ell}$ distribution of the selected candidates, compared to the background expectation for \subref{fig:sub_zoom_range} the $100-250\:\gev$ mass range and \subref{fig:sub_full_range} the
  full mass range of the analysis. Error bars represent 68.3\% central confidence intervals. The signal expectation for several
  $m_{H}$ hypotheses is also shown. The resolution of the reconstructed Higgs mass is dominated by  detector resolution at low $m_{H}$ values and by the 
  Higgs boson width at high $m_{H}$.\label{fig:finalMassesSignal}}
\end{figure*} 

The $Z+\rm{light~jets}$ and $Zb\bar{b}$ backgrounds are evaluated using data. Systematic uncertainties of 45\% and 40\%, respectively, 
are assigned to their normalization to account for the statistical uncertainty in the yield of the control sample, 
the uncertainty in the composition of the control sample, and the uncertainty in the MC-based extrapolation to the signal region. 

The overall uncertainty in the integrated luminosity for the complete 2011 dataset is 3.9\%, 
based on the calibration described in Refs.~\cite{Aad:2011dr,lumi2011} including an additional 
uncertainty for the extrapolation to the later data-taking period with higher instantaneous luminosity.

\section{Results}
In total, \candtotal candidate events are selected by the analysis: \candmu $4\mu$, \candmix $2e2\mu$, and \cande $4e$ events.  
From the background processes, \bkgexp events are expected:  \candmuexp $4\mu$, \candmixexp $2e2\mu$ and \candeexp $4e$. 
In Table~\ref{tab:EventYields}, the number of events observed in each final state is summarized and compared to the expected backgrounds, 
separately for $m_{4\ell}<180\:\gev$ and $m_{4\ell}\geq 180\:\gev$, and to the expected signal for various $m_{H}$ hypotheses. 
The $m_{12}$~and $m_{34}$ mass spectra are shown in Fig.~\ref{fig:finalMasses}. 
The expected $m_{4\ell}$ distributions for the total background and several signal hypotheses are compared to the data in Fig.~\ref{fig:finalMassesSignal}. 

Upper limits are set on the Higgs boson production cross section at 95\% CL, using the $CL_s$ modified frequentist formalism~\cite{cls} 
with the profile likelihood ratio test statistic~\cite{Cowan:2010js}. The test statistic is evaluated with a binned maximum-likelihood fit of 
signal and background models to the observed $m_{4\ell}$ distribution. Figure~\ref{fig:exclusion} shows the observed and expected 
95\% CL cross section upper limits, calculated using ensembles of simulated pseudo-experiments, as a function of $m_{H}$. The SM Higgs boson is excluded at 95\% CL in the mass ranges \excludedranges. 
The expected exclusion ranges are \excludedrangesexp.

The significance of an excess is given by the probability, \pval, that a background-only experiment is more signal-like than that observed. 
In Fig.~\ref{fig:p0} the \pval-values, calculated using an ensemble of simulated pseudo-experiments, are given as a function of \mH\ for the full mass range of the analysis. 
The most significant upward deviations from the background-only hypothesis are observed for 
$m_{H}=\pvaluelowm$ with a local \pval\ of \pvaluelow\ (\sigmalow~standard deviations), $m_{H}=\pvaluehighm$ with a local \pval\ of \pvaluehigh\ (\sigmahigh~standard
deviations) and $m_{H}=\pvaluehighmtwo$ with a local \pval\ of \pvaluehightwo\ (\sigmahightwo~standard deviations). 
The median expected local \pval\ in the presence of a SM Higgs boson are 
\pvaluelowexp\ (\sigmalowexp~standard deviations), \pvaluehighexp\ (\sigmahighexp~standard deviations) and \pvaluehightwoexp\ (\sigmahightwoexp~standard deviations) for 
$m_{H}~=~\pvaluelowm,~\pvaluehighm~\rm{and}~\pvaluehighmtwo$, respectively. 
An alternative calculation, using the asymptotic approximation of Ref.~\cite{Cowan:2010js}, yielded compatible results ---~within 0.2 standard deviations~--- in the entire mass range.

The quoted values do not account for the so-called look-elsewhere effect, which takes into account that such an excess (or a larger one) can appear anywhere 
in the search range as a result of an upward fluctuation of the background.
When considering the complete mass range of this search, using the method of Ref.~\cite{Gross:2010qma}, 
the global \pval-value for each of the three excesses becomes of $O(50\%)$. 
Thus, once the look-elsewhere effect is considered, none of the observed local excesses are significant.

\section{Summary}
A search for the SM Higgs boson in the decay channel \htollllbrief~based on \lumiAverage~ of data recorded by the ATLAS detector at $\sqrt{s}=7\:\tev$ during the 2011 run has been presented. 
The SM Higgs boson is excluded at 95\% CL in the mass ranges \excludedranges. 
The largest upward deviations from the background-only hypothesis are observed for $m_{H}=\pvaluelowm,\pvaluehighm\text{ and }\pvaluehighmtwo$ with local significances of \sigmalow, \sigmahigh~and
\sigmahightwo~standard deviations, respectively. Once the look-elsewhere effect is considered, none of these excesses are significant.

\begin{figure}[!htb]
\centering  
\includegraphics[width=0.99\linewidth]{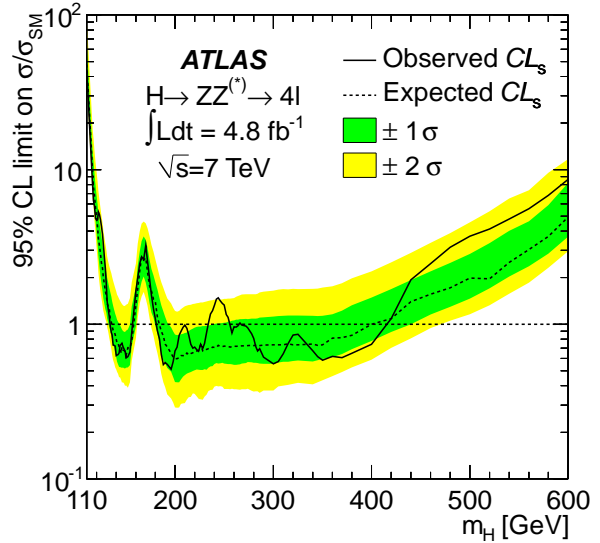}
\caption{The expected (dashed) and observed (full line) 95\% CL upper limits on the Standard Model Higgs boson production cross section as a function of \mH, 
 divided by the expected SM Higgs boson cross section. The dark (green) and light (yellow) bands indicate the 
 expected limits with $\pm 1\sigma$ and $\pm 2\sigma$ fluctuations, respectively.\label{fig:exclusion}}
\end{figure}

\begin{figure}[!htb]
\centering  
\includegraphics[width=0.99\linewidth]{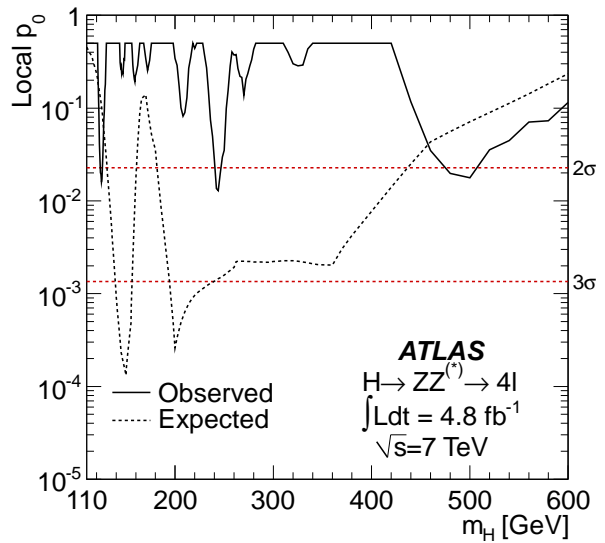}
\caption{
The observed local \pval, the probability that the background fluctuates to the observed number of events or higher, is shown as the solid line.
The dashed curve shows the expected median local \pval~for the signal hypothesis when tested at $m_H$.
The two horizontal dashed lines indicate the \pval~values corresponding to local significances of 2$\sigma$ and 3$\sigma$.\label{fig:p0}}
\end{figure}

\section{Acknowledgements}

We thank CERN for the very successful operation of the LHC, as well as the
support staff from our institutions without whom ATLAS could not be
operated efficiently.

We acknowledge the support of ANPCyT, Argentina; YerPhI, Armenia; ARC,
Australia; BMWF, Austria; ANAS, Azerbaijan; SSTC, Belarus; CNPq and FAPESP,
Brazil; NSERC, NRC and CFI, Canada; CERN; CONICYT, Chile; CAS, MOST and NSFC,
China; COLCIENCIAS, Colombia; MSMT CR, MPO CR and VSC CR, Czech Republic;
DNRF, DNSRC and Lundbeck Foundation, Denmark; ARTEMIS and ERC, European Union;
IN2P3-CNRS, CEA-DSM/IRFU, France; GNAS, Georgia; BMBF, DFG, HGF, MPG and AvH
Foundation, Germany; GSRT, Greece; ISF, MINERVA, GIF, DIP and Benoziyo Center,
Israel; INFN, Italy; MEXT and JSPS, Japan; CNRST, Morocco; FOM and NWO,
Netherlands; RCN, Norway; MNiSW, Poland; GRICES and FCT, Portugal; MERYS
(MECTS), Romania; MES of Russia and ROSATOM, Russian Federation; JINR; MSTD,
Serbia; MSSR, Slovakia; ARRS and MVZT, Slovenia; DST/NRF, South Africa;
MICINN, Spain; SRC and Wallenberg Foundation, Sweden; SER, SNSF and Cantons of
Bern and Geneva, Switzerland; NSC, Taiwan; TAEK, Turkey; STFC, the Royal
Society and Leverhulme Trust, United Kingdom; DOE and NSF, United States of
America.

The crucial computing support from all WLCG partners is acknowledged
gratefully, in particular from CERN and the ATLAS Tier-1 facilities at
TRIUMF (Canada), NDGF (Denmark, Norway, Sweden), CC-IN2P3 (France),
KIT/GridKA (Germany), INFN-CNAF (Italy), NL-T1 (Netherlands), PIC (Spain),
ASGC (Taiwan), RAL (UK) and BNL (USA) and in the Tier-2 facilities
worldwide.

\bibliographystyle{elsarticle-num}
\bibliography{H4lpaper}

\clearpage 
\onecolumn
\begin{flushleft}
{\Large The ATLAS Collaboration}

\bigskip

G.~Aad$^{\rm 48}$,
B.~Abbott$^{\rm 110}$,
J.~Abdallah$^{\rm 11}$,
S.~Abdel~Khalek$^{\rm 114}$,
A.A.~Abdelalim$^{\rm 49}$,
A.~Abdesselam$^{\rm 117}$,
O.~Abdinov$^{\rm 10}$,
B.~Abi$^{\rm 111}$,
M.~Abolins$^{\rm 87}$,
O.S.~AbouZeid$^{\rm 157}$
% ...
\bigskip

% ...
$^{10}$ Institute of Physics, Azerbaijan Academy of Sciences, Baku, Azerbaijan\\
$^{11}$ Institut de F\'isica d'Altes Energies and Departament de F\'isica de la Universitat Aut\`onoma  de Barcelona and ICREA, Barcelona, Spain\\
% ...
$^{48}$ Fakult\"{a}t f\"{u}r Mathematik und Physik, Albert-Ludwigs-Universit\"{a}t, Freiburg i.Br., Germany\\
$^{49}$ Section de Physique, Universit\'e de Gen\`eve, Geneva, Switzerland\\
% ...
$^{87}$ Department of Physics and Astronomy, Michigan State University, East Lansing MI, United States of America\\
% ...
$^{110}$ Homer L. Dodge Department of Physics and Astronomy, University of Oklahoma, Norman OK, United States of America\\
$^{111}$ Department of Physics, Oklahoma State University, Stillwater OK, United States of America\\
% ...
$^{114}$ LAL, Univ. Paris-Sud and CNRS/IN2P3, Orsay, France\\
% ...
$^{117}$ Department of Physics, Oxford University, Oxford, United Kingdom\\
% ...
$^{157}$ Department of Physics, University of Toronto, Toronto ON, Canada\\
% ...
\end{flushleft}


\end{document}
\bye